\documentstyle[11pt,newpasp,twoside,epsf]{article}
\def\edcomment#1{\iffalse\marginpar{\raggedright\sl#1\/}\else\relax\fi}
\reversemarginpar

\begin{document}
\title{Conference Impression}
 \author{Hugo E. Schwarz}
\affil{Nordic Optical Telescope, Apartado 474, E-38700, Sta~Cruz~de~La~Palma, 
Canaries, Spain}

\begin{abstract}
This paper gives a personal impression of the
conference ``Asymmetrical PNe II: From Origins to Microstructures'',
flags some of the highlights, gathers together some facts and
terminology, and indicates some promising future lines of work in this field.
\end{abstract}

\section{Introduction}

These conference proceedings are the condensed result of the 2nd
meeting on Asymmetrical PNe, held at MIT during August of 1999. Being
the last such meeting of this millennium, and a successor of the
successful 1st asymmetrical meeting held at Oranim in Israel during
August 1994, and published as Anns.Isr.Phys.Soc.,11 in 1995, it was
especially good to see that the subject has made significant progress
over the last lustrum. The main impact, as you can see immediately by
just a quick skim through this volume, has come from the imaging done
by the Hubble Space Telescope. The wealth of complex detail revealed
by these unprecedented images has left the theorists reeling, the
observers gasping, and both groups with plenty ground--based work
ahead of them. A special highlight of the meeting was the free
hand--out to participants of a CD with superb WFPC2 images by Hajian
\& Terzian! Note that references to papers in this volume are from here on 
indicated by an asterisk next to the authors name.

\section{Overall impression}

My overall impression from this meeting is that the observed
morphologies of PNe are shown to be more and more complex, mainly due
to the images from HST. The circle with a dot in the middle of 15
years ago is no more! The impact of the increased spatial resolution
data available from the space observations is extraordinary! Archival
research with HST data is now becoming a important tool for PNe
research, especially in combination with ground--based high resolution
spectroscopy. To date there are about 125 HST images of PNe available.
Building on the basis created by the work of Josef Solf in the early
eighties (e.g. Solf 1983, 1984), the spatial resolution and
spectroscopic separation at the sub--pixel level of small features is
now coming of age, and is teaching us a lot about the formation of
PNe. All this detail, however, is clearly posing a problem for the
theorists: the complexity is enormous and at present theory
understandably lags the observations. \\

Having said this, theory has also made progress; several models are
getting better at producing the observed gross shapes of nebulae
(Frank, Garc\'{i}a--Segura, Livio, Soker and others (*)), and
attempts at explaining small scale structure (Dwarkadas(*)) are being
made. The overall theoretical picture, however, is still quite
elusive.  Mass loading obviously plays an important role in some
objects (e.g. globules in the Ring nebula (Dyson(*)), and perhaps also
in NGC6369, and NGC6751 just from their complex morphology, and the
theory for this phenomenon is well developed. \\

Point--symmetry has become very popular and it seems that there are
hardly any asymmetrical PNe left that do not have some kind of
point--symmetry. I recall that at the meeting in La Serena (Mass Loss
on the AGB and Beyond) held in 1992, the concept was hardly known.\\

The use of movies to show time dependent behaviour of data is also on
the increase with several presentations both of observational and
theoretical material at the meeting, giving some interesting new ways
of visualising data. The movie of M2--9's inner nebula (Balick(*))
especially struck me, since it showed the nebula's rotation, which
would have got lost had the data been presented any other way. To
``see'' the ISM penetrating a PN (Villaver(*)) and thus creating an
asymmetry was also spectacular. Internet access to this kind of movie
is also an excellent way of showing the community what is happening.

\section{Terminology}

As a result of the increasing complexity of the observed morphologies,
the terminology has recently also proliferated. Table~1 lists the
terms I heard at this meeting to describe features in PNe. They range
from the practical to the sublime, and even beyond...\\

\begin{table}
\caption{Terms used at this conference to describe morphological 
features in PNe.}
\vspace{0.4cm}
\begin{tabular}{c|c|c|c|c|c|c|c}
\tableline

Jets & FLIERS & Rings & Cocoons & Toroidal cocoons \\
Disks & Ansae & Shells & Trails & Parallelograms   \\
Haloes & Real haloes & Blobs & Lobes & Bullets \\
Tori & Structures & Knots & Filaments & Bubbles \\
Forks & Waists & Hoops & Loops & Double Bubbles \\
Tips & Cylinders & Envelopes & Funnels & Tuning forks \\
Stripes & Dark lanes & Arcs & Caps & Capped bicones\\
Tails & Remnants & Globules & Cometary... & Nipples \\
Rays & Corkscrews & Rolls & Ellipsoids & Shadow columns \\
LIS & Lattices & Petals & Slobs & Smoking guns \\
\tableline\tableline
\end{tabular}
\end{table}

``Paradigm'' is definitely in, the previously much over-used
``scenario'' is out; ``Weather'' (which you ignore...) is in;
``Physics'' (...which you don't) is out. Other much used key words
were: precession, point--symmetry, BRETS, episodic or periodic
outflows, irregular, mass loading. Finally, a nice piece of
nomenclature was reportedly found in the address of a famous
Observatory, the ``Navel Absorbatory''.\\

It is clear that astronomers are also influenced by "fashions" or
"trends". Please start a trend and use the word ``polarigenic''
(coined by yours truly about 15 yrs ago), as in "polarigenic
mechanism" for something that produces polarisation.\\

\section{Observational appearance of bipolar and point--symmetric PNe.}

Figure~1 shows the possible appearances of bipolar and
point--symmetric nebulae as projected onto the plane of the sky. Under
certain circumstances blue and red shifted components can be located
on the same side of the object, even though this is, at first sight,
counter intuitive. For objects with a precession cone cut by the line
of sight or plane of the sky this is observed (e.g. IC4634) and such
objects are point--symmetric. M2--9 is a bipolar that shows plane
symmetry in it's inner nebula, and point--symmetry in it's outer lobes
which are also both red shifted since they are reflecting light from
the central object. So far this is unique among PNe. Nebulae are put
into the following classes:

Bipolars: NO (near, or not too far from plane of sky), EO (nearly 
end--on), RR (both lobes red shifted)

Point--symmetricals: P1 ( 0 $\geq$ i $\leq$ 0.5$\phi$) , P2 (0.5$\phi$
$\geq$ i $\leq$ 90-0.5$\phi$), P3 (90-0.5$\phi$ $\geq$ i $\leq$ 90),
where $\phi$ the precession cone opening angle, and i is the
inclination angle to the plane of the sky of the nebula.

\begin{figure}
\plotone{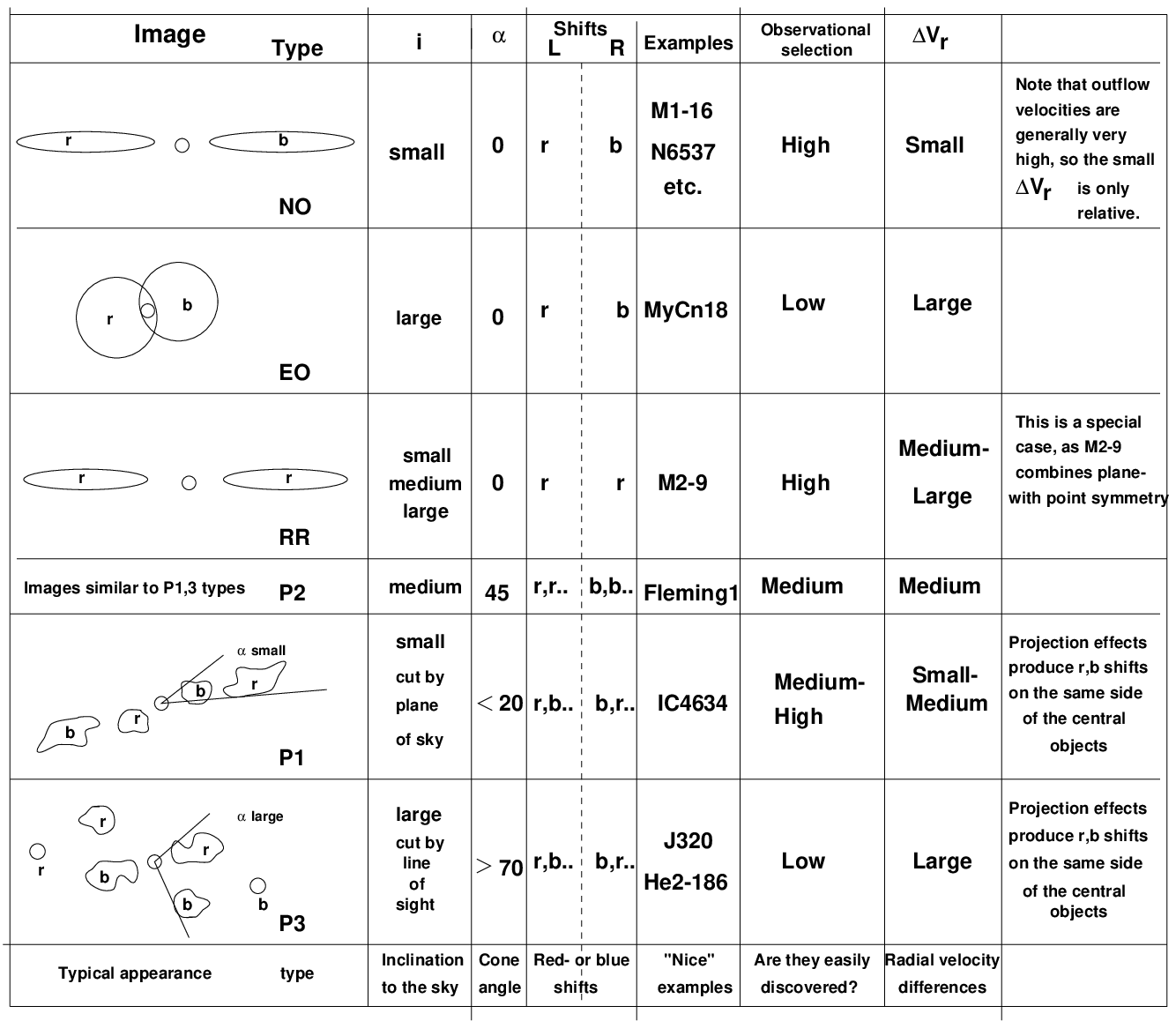} 

\caption{The possible appearances of bipolar and
point--symmetric nebulae are shown. The inclinations are assumed and
result in these appearances. The ease of discovery of bipolar nebulae
depends on their inclination: near the plane of the sky they present a
bipolar aspect which makes (typically) imaging discovery easier. The
opening angle for precession cones is taken to be smaller than
40$^o$, and $\alpha$ is the projected cone angle. For P2 nebulae, the
precession cone is not cut by the plane of the sky or the line of
sight, and this type does {\it not} show blue and red shifted
components on the same side of the central object. Note that type EO
can also have B and R shifts on the same side of the object, and that
the radial velocity difference refers to the projected blue and red
lobe velocities.}
\end{figure}

\section{Future work}

Some lines of research are promising and should be continued in the
near future. Here I mention a few, based on what I heard and saw at
the meeting, that I find of particular potential interest.

\subsection{Binaries versus single stars}

The whole issue of what kind of mechanism drives the production of the
most asymmetrical and physically different class of PNe, the bipolar
nebulae (Corradi \& Schwarz, 1995), depends critically on finding the
binary central stars. There is strong circumstantial evidence for the
presence of binaries in several bipolars, but it has so far been very
difficult to obtain direct determinations of the numbers of bipolars
and other PNe that contain binary central stars. A promising method is
that of radial velocity determinations, especially in the infra--red,
where obscuration is less of a problem. Bond (*) has done pioneering
imaging work with HST, which is rightly being continued.\\

From the theorist's point of view the single versus binary star model
for bipolar nebulae is split into two parts: the making of a bipolar
nebula starting with an asymmetrical density distribution, and the
mechanism to produce that initial density distribution in the first
place. To make a bipolar nebula with some outflow in an asymmetrical
density distribution, is relatively easy and many models can do
this. Particularly successful are the models of Garc\'{i}a--Segura
(*), a two wind interaction combined with magnetic field and near
break--up rotation, and the work by Frank et al. (*) based originally
on work by Icke (1988) using the Kompaneets theory. The nebulae with
the highest aspect ratios, such as M2--9, pose more of a problem as
the required density gradient and other parameters become
un--physically high. Here the binaries form a more natural system to
generate a strong pole to equator density gradient, and their
accretion disks can blow disk winds that are highly collimated and
fast, explaining the high observed velocities, the strong collimation,
and the large wings of the observed hydrogen line profiles. Excretion
disks, as proposed by Morris (1987), can shape the slow wind from the
primary star into two lobes, through which the fast wind then blows
out HH--like objects as in He2-104. In single stars there is no
identified mechanism to produce the necessary high underlying density
gradient that shapes the wind, and a particular combination of strong
magnetic fields, rotation, and equatorial mass loss has to be invoked
to give asymmetrical nebulae at all. My overall impression is that
binaries are needed for the most extreme bipolars and perhaps for all
bipolars. Point--symmetry is most easily explained by precession;
again something that occurs naturally in binaries. Also the
combination in M2--9 of both point-- and plane symmetry is difficult
to explain without a binary central object, as is the presence of
[OIII] lines and a low luminosity central object. I think that an
object like He3--1475 (see the HST image on the CD handed out at the
meeting) cannot be explained by any single star model! We must go and
find those binaries!

\subsection{PNe--ISM interactions}

There were several papers on the interactions of PNe with the ISM
(Knill--Degani; theory, Kerber; observations, Villaver; modeling, with
movie(*)). This is clearly a very interesting asymmetry producing
mechanism, especially affecting the outer parts of the PNe, and in
some cases even resulting in stars being outside their own nebula. The
ratio of density between a typical PN and the ISM is about 50 in the
Galactic plane and 300 in the halo. Under certain circumstances the
ISM can penetrate the PN, giving spectacular interactions, as shown in
the presented movie.

\subsection{Morphology in the HR diagramme}

The morphologies derived from HST images of a sample of PNe taken and to
be taken in the Magellanic Clouds, can yield important relationships
between morphology and the central star evolutionary stage. This work was
started by using Galactic PNe, extracting statistical information
about a sample of which narrow band images had been taken, and making
a morphological classification of the nebulae. The result was a plot
of their central stars in the HR diagram as a function of
morphological class, yielding some intriguing correlations, especially
for the bipolars (Stanghellini et al.1993). The large uncertainty in
the individual distances to the PNe in this sample make this result
only statistically interesting. By going to the Clouds, the distance
uncertainty is gone and much better results can be obtained, again
using HST. Stanghellini is undertaking this important work.

\subsection{Polarimetry}

Polarimetry was meagrely represented at this conference, as at
most meetings that do not specifically deal with the subject. Only one
out of 55 talks and 2 out of 38 posters contained some item about
polarimetry. Since polarimetry often can provide a missing and vital
piece of information, this is a plea for all observers to use and be
more aware of polarimetry. The fact that the title of this meeting
contains the word "asymmetrical" should already make people think
about polarimetry, as it is uniquely suitable to detect
asymmetries. \\

Two examples of the importance of polarimetry are: Trammell et
al. (1994) found that 77\% of 31 AGB objects are intrinsically
polarised, while Johnson and Jones (1991) found 74\% of their sample
to be polarised. There clearly is a high fraction of AGB stars with
asymmetries. If there are no features in the polarisation spectrum,
there has to be a global asymmetry in the object, and probably dust
scattering; if there are spectral features, local asymmetries such as
convection cells or other atmospheric features must be present. The
whole critical question of the onset of asymmetry during late stellar
evolution, whereby essentially spherical stars produce PNe of which
many are asymmetrical, can be uniquely studied with polarimetry.  The
mechanism of dust scattering in the faint loops in M2-9 (Schwarz et
al. 1997) was a supposition until the loops were found to be 60\%
polarised at right angles to the plane formed by the source, scatterer
and observer. Such high degree of polarisation is not only easy to
observe it also firms up the proposed dust scattering mechanism to a
certainty. This then allowed the distance to M2--9 to be determined
accurately from it's expansion parallax. Polarimetry provided the key
piece of information that allowed the physical parameters of this
object to be found.\\

Continued and increased work using polarisation measurements on PNe
and AGB stars--such as that in the press (ApJ) by Weintraub, Kastner
(who showed the data at this meeting), Sahai, \& Hines, who used
NICMOS to image AFGL2688--will surely yield important results! \\

In summary, this was a meeting clearly dominated by the observers
using HST imagery making theorist's lives difficult. The old Chinese
curse ``May you live in interesting times!'' is applicable here, and
times will become even more interesting in the near future. Thanks for
inviting me, Joel and Noam, and see you at the next asymmetrical
meeting!

\end{document}